\documentclass[prl,twocolumn,fleqn]{revtex4}
\usepackage{amscd,verbatim}
\usepackage[all]{xy}
\usepackage{graphicx}

\usepackage{amsmath}
\usepackage[psamsfonts]{amssymb}
\usepackage{mathtools}
\usepackage{euscript}

\usepackage{latexsym}

\usepackage{enumerate}

\renewcommand{\(}{\begin{equation}}
\renewcommand{\)}{end{equation} \vspace{-.05in}\linebreak}

\newcounter{saveeqn}
\newcounter{savealpheqn}

\newcommand{\alpheqn}{\setcounter{saveeqn}{\value{equation}}%
  \stepcounter{saveeqn}\setcounter{equation}{0}%
  \renewcommand{\theequation}{\mbox{\arabic{section}.\arabic{saveeqn}
\alph{equation}}}
  \renewcommand{\)}{\end{equation}}}
\def\part#1{\frac{\partial}{\partial{#1}}}%
\def\group#1{\refstepcounter{equation}\setcounter{saveeqn}
 {\value{equation}}%
  \label{#1}\setcounter{equation}{0}%
\renewcommand{\theequation}{\mbox{\arabic{section}.\arabic{saveeqn}
\alph{equation}}}
  \renewcommand{\)}{\end{equation}}}
\newcommand{\reseteqn}{\setcounter{equation}{\value{saveeqn}}%
  \renewcommand{\theequation}{\arabic{section}.\arabic{equation}}%
  \renewcommand{\)}{\end{equation}}}

\newcommand{\aalpheqn}{\setcounter{saveeqn}{\value{equation}}%
  \stepcounter{saveeqn}\setcounter{equation}{0}%
  \renewcommand{\theequation}{\mbox{
        \Alph{subsection}.\arabic{saveeqn}\alph{equation}}}
   \renewcommand{\)}{\end{equation}}}
\newcommand{\areseteqn}{\setcounter{equation}{\value{saveeqn}}%
  \renewcommand{\theequation}{\Alph{subsection}.\arabic{equation}}%
  \renewcommand{\)}{\end{equation}}}

\renewcommand{\thefootnote}{\alph{footnote}}
\renewcommand{\(}{\begin{equation}}
\renewcommand{\)}{\end{equation}}
\newcommand{\ba}{\begin{eqnarray}}
\newcommand{\ea}{\end{eqnarray}}

\newcommand{\bp}{\mathop{\vtop{\ialign{##\crcr
   $\hfil\displaystyle{}\hfil$\crcr\noalign{\kern-13pt\nointerlineskip}
   \BIG{(}\hskip0pt\crcr\noalign{\kern3pt}}}}}
\newcommand{\cbp}{\mathop{\vtop{\ialign{##\crcr
   $\hfil\displaystyle{}\hfil$\crcr\noalign{\kern-13pt\nointerlineskip}
   \BIG{)}\hskip0pt\crcr\noalign{\kern3pt}}}}}
\newcommand{\pa}{\mathop{\vtop{\ialign{##\crcr

$\hfil\displaystyle{\oplus}\hfil$\crcr\noalign{\kern+1pt\nointerlineskip
}
   \hspace{.08in}$^{\alpha=0}$\hskip6pt\crcr\noalign{\kern3pt}}}}}

\newcommand{\cF}{\ensuremath{\mathcal F}}

\newcommand{\Q}{\ensuremath{\mathbb Q}}

\newcommand{\Z}{\ensuremath{\mathbb Z}}

\def\cO{\ensuremath{{\cal O}}}

\newcommand{\beq}{\begin{equation}}
\newcommand{\eeq}{\end{equation}}

\newcommand{\Or}{{\rm O}}
\newcommand{\Spin}{{\rm Spin}}
\newcommand{\String}{{\rm String}}
\newcommand{\Five}{{\rm Fivebrane}}
\newcommand{\Nine}{{\rm Ninebrane}}

\numberwithin{equation}{section}
\renewcommand{\theequation}{\mbox{\arabic{equation}}}

\catcode`\@=11
\def\vereq#1#2{\lower3pt\vbox{\baselineskip1.5pt \lineskip1.5pt
\ialign{$\m@th#1\hfill##\hfil$\crcr#2\crcr\sim\crcr}}}
\catcode`\@=12

\makeatletter
\newcommand\figcaption{\def\@captype{figure}\caption}
\newcommand\tabcaption{\def\@captype{table}\caption}
\makeatother
\renewcommand{\(}{\begin{equation}}
\renewcommand{\)}{\end{equation}}

\newcommand{\CC}{{\mathbb C}}

\begin{document}
\def\thefootnote{\fnsymbol{footnote}}

\title{Topological actions via gauge variations of 
higher structures}

\author{Hisham Sati$^1$ and Matthew Wheeler$^{2}$}
\affiliation{
$^1$ Division of Science and Mathematics, New York University, Abu Dhabi, UAE.\\
$^2$ Department of Mathematics, The University of Arizona,Tucson, AZ 85721, USA.}

\begin{abstract}
In this note we provide a new perspective on the topological parts of several action functionals in string 
and M-theory. We show that rationally these can be viewed as large gauge transformations corresponding to 
variations of higher structures, such as String, Fivebrane, and Ninebrane 
structures. 
\end{abstract}

%
\setcounter{footnote}{0}
\renewcommand{\thefootnote}{\arabic{footnote}}


\maketitle

\renewcommand{\thepage}{\arabic{page}}

\section{Introduction}

Dirac's formulation of spinors establishes that the tangent space with its rotation group (in the Euclidean
setting) or Lorentz group (in the Minkowski case) is not enough to account for their behavior and properties.
The proper formulation is via lifting to the Spin group, which is the double cover of the orthogonal group.
From a topological point of view, one requires that the spacetime manifold $M^n$ admit a Spin structure, 
whose existence can be established at the level of classifying space ${\rm BSpin}(n)$ 
and requires the vanishing of the second Stiefel-Whitney class $w_2$ of $M^n$. Had the Spin group 
not already been known, one could have further defined the Spin group to be an appropriate loop space 
of  BSpin$(n)$.

 Generalizing from point particles to strings, one realizes that a Spin structure is no longer
able to  fully account for anomalies, and an appropriate formulation requires lifting to a String structure  
\cite{Kil}. This still allows for using Dirac operators and to extending to other backgrounds,
such as orbifolds \cite{PW}. To account for other signatures, including Lorentzian, semi-Riemannian 
analogues of String structures are constructed in \cite{SS}. 
From the point of view of (higher) groups, one has the String group as the loop space of the classifying 
space BString$(n)$ of such structures. 
 Similar, but more subtle, 
 arguments hold in part for the generalization to the fivebrane \cite{SSS2}\cite{SSS3}  
 and the ninebrane \cite{9brane}, where one can also lift to groups (in a more general homotopy sense)
 that are loop spaces  of the corresponding classifying spaces BFivebrane$(n)$ and BNinebrane$(n)$, respectively.
 The structures are related via the Pontrjagin classes 
$$
\label{seq}
\xymatrix@=1.3em{
\text{\small SO}   \ar@{~>}[r]  &   \text{\small Spin}   \ar@{~>}[r]^-{p_1^\Q} \; &
\text{\small String}  \ar@{~>}[r]^-{p_2^\Q} \; & \text{\small Fivebrane}
 \ar@{~>}[r]^-{p_3^\Q} \; & \text{\small Ninebrane}.  } 
 $$
 The above structures are in the spirit of the higher algebraic approaches to M-branes 
\cite{BL}\cite{Gu}\cite{ABJM}, but more on the topological side. By uncovering 
such structures (see \cite{T} for another recent illustration), one hopes that more insight is gained into the 
fundamental nature of M-theory and its relation to string theory. 
A priori there are several subtleties involved in considering the above higher
 structures; most notably the presence of torsion in cohomology and the need to use higher
generalization of bundles, i.e., higher gerbes or $n$-bundles. While such formulations have been 
applied in useful ways, e.g., to the worldvolume theories of M-branes 
(see \cite{tcu}\cite{mem}\cite{Sa5}\cite{FSS1}\cite{SSc} for various approaches),
both of these complications can be set aside 
by rationalizing the structures, i.e., by taking the corresponding cohomology to be  
over the rational (or real) numbers. The outcome is a resolution of both subtleties with the same token:
not only does torsion get evaded, but also the higher structures themselves can now be 
described using a formulation via only the much more familiar Spin structures and corresponding bundles,
which after all is sufficient for many purposes in physics.

Indeed, in \cite{SW1} we defined and characterized these rational Spin-String, Spin-Fivebrane,
 and Spin-Ninebrane  structures. It is our aim here to apply these  to describe how topological action functionals in 
M-theory and string theory can be interpreted as  global gauge transformations corresponding 
to variations of such structures. Such a point of view on the actions via variations of structures has been 
proposed in \cite{top}, where  variations of framing 
(essentially a parallelism, i.e. with a trivialization of the tangent bundle) was highlighted. 
Our current description is a generalization to the case when the structure is topologically 
(highly) nontrivial.

 We consider systems arising from string theory and M-theory via action functionals
which, to some extent, allow us to classically describe the dynamics of the system. 
However, we are interested in considering the ``topological terms" which are independent 
of the metric on the underlying manifold. These terms have the virtue that
they are trusted to some extent upon probing the quantum regime. Therefore, focusing 
on the topological terms  allows for at least setting up a starting point for 
a corresponding  quantum field theoretic construction, namely the partition function $Z$
 (see \cite{Fr}\cite{FM}\cite{SSS2} for detailed treatments associated with such a problem).
It is natural to ask whether the physical entities, e.g. the action functional
and the partition function, depend on the underlying geometric and 
topological structures imposed on spacetime. For instance, (in)dependence
on the underlying smooth structure plays an important role in 
global anomalies \cite{Wi}. Similarly, dependence on the Spin 
structure is a classical question in string theory \cite{SW}. 
One can also ask a related question: How do the physical entities
behave under the variation of a corresponding structure? 
While we do not attempt a full and general treatment here, 
we do describe the connection in the M-theory and string theory setting: 
  for the case of 
 Spin-Fivebrane structures in relation to the NS5/M5-brane action, and then  
  for Spin-Ninebrane structures for the Green-Schwarz anomaly and its dual as well as for 
  the Chern-Simons and one-loop terms in M-theory. 
  
  \medskip
  Interpreting the cup products from \cite{SW1} as topological Lagrangians for brane 
sigma models, means to regard these branes as propagating not on spacetime M, but on the
total space of the Spin bundle $Q$, i.e., we have sigma models on spacetimes which are 
``extended".  Indeed, in \cite{DS},  the authors describe the heterotic string not as an 
algebraically defined CFT, but as a geometric sigma model  on the Spin bundle $Q$ of spacetime 
$M$. In this   ``fibered $(0,2)$ WZW model'',  the Green-Schwarz-twisted $B$-field is viewed 
  as a bundle gerbe on $Q$ which restricts on each fiber to the canonical bundle gerbe 
  on Spin, hence to the topological term for the fiberwise WZW model. This means, in particular, 
  that the degree three class of the bundle gerbe on Q restricts on each fiber to the canonical 3-form 
  on Spin. We interpret this here as the String structure   $\mathcal{S}$ (see below). 
  
\medskip
Since the action is fiberwise a WZW model, these additional fiberwise degrees of freedom are 
exactly what is expected to gives the extra current algebra degrees of freedom seen in the heterotic string, 
but not seen in an ordinary geometric sigma-model on spacetime $M$ \cite{DS}.
Our approach then can be viewed as extending this case to other situations in M-theory and string theory, 
namely for the M-theory action and for the NS5/M5-brane action, albeit at the rational level.  
On the other hand, and in a dual sense,  here we are looking at  the 
higher brane version of Distler-Sharpe's heterotic string sigma models on $Q$ (instead of on $M$) \cite{DS}, 
as an analogous ``fibered'' version of the heterotic NS5-brane sigma-model.
Such a heterotic perspective  fits well with the general setup as Fivebrane structures \cite{SSS2}\cite{SSS3}
are motivated from anomalies seen specifically for the heterotic NS5-brane in the ``magnetic dual" 
formulation of heterotic string theory. Other instances of extended spacetime can be found in \cite{T}
and references therein. 


%

\section{Variations of structures as global gauge transformations} 
\label{Sec App}
%

Generally, insisting that the partition function $Z$ of the topological part of the 
action functional $S$ be defined and well-defined as a function on the moduli 
space of fields imposes conditions on the underlying manifolds and bundles on
them, in the form of constraints on primary (and possibly secondary) 
characteristic classes. A very classical example, mentioned above, is Dirac's theory of 
fermions, 
which requires a Spin bundle lift of the
 tangent bundle, i.e.  the underlying spacetime 
$M$ has to have a vanishing second Stiefel-Whitney class, 
$w_2(M)=0$. Other instances arise in string theory 
and M-theory where a lift to one of the higher connected covers 
discussed earlier is needed (see \cite{SSS2}\cite{SSS3}\cite{9brane}).

It is natural to ask whether the physical entities, e.g. $S$ or $Z$, 
depend on the underlying geometric and 
topological structures on spacetime: How do the physical entities
behave under the variation of a corresponding structure? 
For instance, for Chern-Simons theory, the variation of the
Spin structure was 
considered in \cite{J}, where two Spin structures essentially differ
by some real line bundle. Hence once $w_2(M)=0$ is satisfied we
seek possible dependence on the elements in $H^1(M; \Z_2)$ or, equivalently,
 real line bundles. 
In going to higher structures, 
in \cite{tcu} dependence of the M2-brane 
partition function on the underlying String structure imposed was studied
via gerbes, as elements in $H^3(M; \Z)$.
Since the M2-brane worldvolume is 3-dimensional, then 
it automatically admits a String structure. The main point in
\cite{tcu} is to identify the action functional as 
 a variation of the String structure, using the constructions in \cite{Red}. 
Here what we really mean by a variation of a $\cO$-structure is 
\(
\label{var}
\text{Variation of $\cO$-structure}=\delta \cO = \cO - \cO'\;,
\)
the difference of two elements in the space of $\cO$-structures. 
When a certain field $\phi$ is identified with a structure $\cO$
(pulled back to spacetime or worldvolume) then we view \eqref{var} as 
a global or large gauge transformation for $\phi$.

In the current context, the space of such structures is the cohomology
group in degree one less than that of the obstruction for the 
given structures, i.e. $H^3(M; \Z)$, $H^7(M; \Z)$, and $H^{11}(M; \Z)$, 
for String, Fivebrane, and Ninebrane structures, respectively. 
We are rationalizing the structures, so that these cohomology groups
take rational coefficients. Furthermore, these `moduli spaces' are 
vector spaces, so that there is  no obstruction to moving in that space, i.e.,
to forming the variations \eqref{var}. We will see situations where the left-hand 
side is the difference between two rational Fivebrane 
structures $\cF - \cF'$ or two rational Ninebrane structures 
$\mathcal{N} - \mathcal{N}'$,
respectively. These  will be given essentially by corresponding lower structures 
on the right-hand sides, namely, a String structure and a Fivebrane 
structure.

Varying underlying topological structures has been advocated in \cite{top} for
 the case of framed manifolds, where
topological parts of the action functionals for the membrane (M2-brane) 
and the fivebrane (M5-brane) worldvolume theories 
were interpreted via a change of framing formula, in the 
spirit of the variational principle. A similar discussion holds for the NS5-brane \cite{NS5}. 
Here we adopt that point of view, leading us to interpret the topological 
action functionals as variations of higher structures. Being in the rational 
setting makes the identification more direct and transparent. 
We emphasize that, while we start with de Rham expressions, we will 
be interested in Lagrangians and action functionals promoted (explicitly or implicitly) to  
the level of cohomology, as in \cite{Wi2}\cite{DFM}\cite{tcu}\cite{top}.
The action functionals involve classes on the total spaces of the principal
bundles, but we are considering expressions on the base spacetime. 
This can be done either by assuming a section, as is the case for
Chern-Simons theory, or via a more elaborate process such as using a Hodge decomposition 
as in \cite{Red}. This will be implicit below and, as such, 
we are not aiming for the most general case.

The constructions and results in \cite{SW1} are established for rational numbers, but they
readily extend over the reals, especially for smooth manifolds as we consider here.
Note that we are considering only topological terms and our expressions 
are cohomological, and so we are in a setting akin to that of a topological field theory.
By variation we mean large gauge transformations that are not expected to hold locally,
hence the parameters do not vary smoothly.

%

\section{Variations on rational Fivebrane classes}
\label{Sec var5}

Here we show that the topological action associated with the NS5-brane and the  M5-brane can be
interpreted via variations of Fivebrane classes. Starting with the String group as a structure group, 
 a $\String$-principal bundle $\pi_{{}_\String}:P\to M$ has a {\rm rational Fivebrane structure} 
if there is a lift of the rationalized classifying map $f:M\to B\String_\Q$ to the homotopy fiber of 
(a fraction)
 of the second rational Pontrjagin class $p_2^\Q$ (note that since we are working rationally, the fractions are inconsequential).
A {\rm rational Fivebrane structure class} is a cohomology class $\mathcal{F}\in H^7(P;\Q)$ such that $\iota_x^*\mathcal{F}=a_7\in H^7(\String;\Q)$ for each fiber inclusion $\iota_x:\String\to P$.
Such higher structures (beyond Spin) have been recast using Spin structures.
 We have shown in \cite{SW1} that  
 
 \vspace{-3mm}
\begin{enumerate}[{\bf (i)}]
\item For every rational Spin-Fivebrane class 
$\mathcal{F}\in H^7(Q;\Q)$, the pullback $\rho^*\mathcal{F}$ is a rational Fivebrane class.

\vspace{-1mm}
\item   For any rational $\Five$ structure $\mathcal{F}\in H^7(P;\Q)$ there is
		 a Spin-Fivebrane class  $\tilde{\mathcal{F}} \in H^7(Q;\Q)$ such that 
		$\rho^*\tilde{\mathcal{F}}=\mathcal{F}$.

\vspace{-1mm}
\item  Two classes $\mathcal{F}$, $\mathcal{F}'\in H^7(Q;\Q)$ give the same Fivebrane 
structure if $\mathcal{F}-\mathcal{F}'=\mathcal{S}\cdot\pi_\Spin^*\phi_4$ where 
$\mathcal{S}\in H^3(Q;\Q)$ is the String structure class and $\phi_4\in H^4(M;\Q)$ 
is a rational cohomology class.
\end{enumerate}

\vspace{-2mm}
\noindent Thus two Fivebrane structures are identified rationally if their difference corresponds to 
a torsion class in $H^7(M;\Z)$. This says that, rationally, all the information on Fivebrane 
structures is essentially encoded in the underlying Spin bundles.

\medskip
{\bf Example 1: The M5/NS5-brane action and variation of Spin-Fivebrane structures.} 
Consider the M5/NS5-brane on an extended 
worldvolume, which is a seven-dimensional Spin manifold $X^7$, as in \cite{W}\cite{Wi3}. 
The action functional of the fivebranes has been considered in 
\cite{tcu}\cite{Sa5}\cite{FSS1} from the point of view of String 
bundles with String connections. At the level of differential forms, the topological part 
is given as 
\(
S_{{}_{{\rm M5}/{\rm NS5}}}=\int_{X^7} C_3 \wedge G_4\;,
\)
where $C_3$ and $G_4$ have the usual M-theory meaning for the 
M5-brane and are different for the NS5-brane (see \cite{Le} where the 
corresponding 6-dimensional term is studied). 
Passing to cohomology, we consider the pairing of corresponding 
cohomology classes with the fundamental homology class of the manifold
\(
S_{{}_{{\rm M5}/{\rm NS5}}}^{\rm coh}=\langle [C_3] \cup [G_4] , [X^7]\rangle\;,
\)
with $G_4=$non-exact $+dC_3$, so that we are able to take both $C_3$ and $G_4$ to be closed
(see \cite{DFM}). 
The heterotic perspective with target $Q$ mentioned in the Introduction and based on \cite{DS} 
 fits naturally here as, due to Green-Schwarz, $C_3 = H_3 + CS_3$
contains a closed 3-form as a summand, without itself being closed, due to the Chern-Simons term.
Indeed, the heterotic NS5-brane action has been considered from a superspace perspective in \cite{Le}. 
Some aspects of this, but in a restrictive situation, extends to M-theory in the flat case, due to 
 the quantization condition 
\cite{Wi2}, or alternatively upon working in heterotic M-theory \cite{HW}.

 Note that the class of $C_3$ or $H_3$ can be interpreted   \cite{tcu} (up to a shift) 
as a String structure $\mathcal{S}$. The integrand then corresponds to 
the class $\mathcal{S} \cup \phi$, with $\phi=[G_4]$ being a rational 
degree four cohomology class. Indeed, $[G_4]$ integrally satisfies the quantization 
condition (see \cite{Wi2})  $G_4 + \tfrac{1}{2}\lambda \in H^4(Y^{11}; \Z)$, where 
$Y^{11}$ is the ambient spacetime into which $X^7$ (and its bounding space)
is mapped. When we rationalize, the requirement that the first 
Spin characteristic class $\lambda=\tfrac{1}{2}p_1$ is divisible 
by two (i.e. $Y^{11}$ admitting a ``Membrane structure'' \cite{mem})
is automatically satisfied, in which case $[G_4] \in H^4(Y^{11}; \Q)$.  Therefore, we 
identify the integrand, i.e., the Lagrangian at the level of cohomology, 
as the difference of two rational Spin-Fivebrane structures on $Y^{11}$. That is, 
\(
\mathcal{L}_{{}_{M5}}^{\rm coh}= \mathcal{F}_\Q - \mathcal{F}'_\Q\;.
\) 
This means that for the M5/NS5-brane action, there exists two Spin-Fivebrane structures such 
that their difference is that part of the Lagrangian.
As in the case of the 3-dimensional M2-brane worldvolume 
admitting a String structure \cite{tcu}, the extended M5/NS5-brane worldvolume 
automatically admits a Fivebrane structure by virtue of it being 7-dimensional. 

Note that  if $H^4(M;\Q)$ is torsion, then the set of Fivebrane classes and Spin-Fivebrane 
classes coincide \cite{SW1}. Compactification manifolds for which 
 this occurs include the following manifolds as in the realistic 
Kaluza-Klein list \cite{Wi-81}.

\medskip
\noindent{\bf Examples:}

\noindent {\bf (i)} The Witten manifolds $M_{k, \ell}$, 
which are $S^1$ bundles over the product of complex projective
spaces $\CC P^2 \times \CC P^1$, are classified in \cite{KS} according to 
two integers $k$ and $\ell$. They have $H^4(M_{k, \ell}; \Z)=\Z/\ell^2$. 

\noindent {\bf (ii)} Generalized Witten manifolds $N_{kl}$ are defined as the total spaces
of fiber bundles with fiber the lens space $L_k(\ell_2, \ell_2)$ 
and structure group $S^1$. They have $H^4(N_{kl}; \Z)\cong \Z_{|\ell_1 \ell_2|}$ \cite{E}. 

\noindent {\bf (iii)} Quaternionic line bundles $E$ over closed Spin manifolds of 
dimension $4k-1$ with $c_2(E) \in H^4(M)$ been torsion considered in 
\cite{CG} via generalizations of the Kreck-Stolz invariants.

\section{Variations on rational Ninebrane classes}
\label{Sec var9}

We now extend the results from the last section to the next  higher connected cover of the 
orthogonal group $\Or$ in the sequence 
 in the Introduction. This allows to   describe the terms in the Green-Schwarz anomaly 
 cancellation and its dual, as well as the M-theory topological terms. 
Note that here things become a bit subtle, as there are two structures sitting in between Fivebrane 
and Ninebrane, denoted 2Orient and $2\Spin$, respectively \cite{9brane}. However, these 
are defined via mod 2 obstructions, so that rationalization will 
make them equivalent to a Fivebrane structure. 
So, to follow along the lines of rational $\Five$ structures, we may define rational $\Nine$ structures  \cite{SW1}.

\vspace{-2mm}
\begin{itemize}
\item A $\Five$-principal bundle $\pi_{{}_{\Five}}:T\to M$ admits a {\rm rational Ninebrane structure} 
if there is a lift of the rational classifying map $f:M\to B\Five_\Q$ to the homotopy fiber 
$F(\tfrac{1}{240}p_3)^\Q$.

\vspace{-1mm}
\item  A {\rm rational Ninebrane class} is a cohomology class $\mathcal{N}_\Q\in H^{11}(T;\Q)$ such 
that $\iota_x^*\mathcal{N}=a_{11}\in H^{11}(\Five;\Q)$ for each inclusion $\iota_x:\Five\to T.$
\end{itemize}

\vspace{-3mm}
\noindent Now, just as we did in the case of Fivebrane structures, we can relate these classes to ones on
the underlying Spin bundle.  In order to do this, as we compared degree 7 rational cohomology between 
Spin and String, we need to compare the degree 11 rational cohomology of Spin and Fivebrane.
 The map $\rho:\Five\to \Spin$ induces an isomorphism 
 $\rho^*:H^{11}(\Spin;\Q)\xrightarrow{\cong}H^{11}(\Five;\Q)$ \cite{SW1}, which  were  
 used  to relate rational Ninebrane classes to classes on the underlying Spin bundle.  
A {\rm rational Spin-Ninebrane class} is then a cohomology class $\mathcal{N}_\Q$ in  $H^{11}(Q;\Q)$ 
such that $\iota_x^*\mathcal{N}_\Q=\tilde a_{11}\in H^{11}(\Spin;\Q)$ for each $x\in M$.
In this case, we have shown in \cite{SW1}:

\vspace{-2mm}
\begin{enumerate}[{\bf (i)}]
\item For every rational Spin-Ninebrane class $\mathcal{N}_\Q\in H^{11}(Q;\Q)$, 
the pullback $\rho^*\mathcal{N}_\Q$ is a rational Ninebrane class.
		
		\vspace{-1mm}
		\item   Any rational Ninebrane structure $\mathcal{N}_\Q\in H^{11}(T;\Q)$ can be described by a class in $H^{11}(Q;\Q)$.
		
		\vspace{-1mm}
		\item   Two classes $\mathcal{N}_\Q, \mathcal{N}'_\Q\in H^{11}(Q;\Q)$ will give the same rational Ninebrane structure if 
		\(
		\label{def-nn}
		\mathcal{N}_\Q- \mathcal{N}'_\Q=\mathcal{S}\cdot\pi_\Spin^*\psi_8+
		\mathcal{F}\cdot\pi_\Spin^*\phi_4\;,
		\)
		 where $\mathcal{S}\in H^3(Q;\Q)$ is the String structure class, $\mathcal{F}\in H^7(Q;\Q)$ is the Fivebrane structure class, while $\psi_8\in H^8(M;\Q)$ and  $\phi_4\in H^4(M;\Q)$ are rational  cohomology classes.
\end{enumerate}

%

\vspace{-2mm}
\noindent We now consider in the  right degree a fundamental example in M-theory, as the topological 
part of 11-dimensional supergravity \cite{CJS} together with the one-loop term \cite{DLM}.
Large gauge transformations in M-theory have been considered from a geometric perspective
in \cite{KaS}, while the full symmetries of the C-field are explained in \cite{DFM}\cite{FSS2}.  

\medskip
{\bf Example 2: The M-theory action and variation of Spin-Ninebrane structures.}
We now consider M-theory on a String manifold $Y^{11}$, as in \cite{E8}. 
The known topological action functional of M-theory is
given as 
\(
\int_{Y^{11}} \left( \tfrac{1}{6}G_4 \wedge G_4 \wedge C_3 - I_8 \wedge C_3   \right)
\label{M action}
\) 
 where $I_8$ is called the one-loop polynomial,
  whose corresponding  cohomology class is $\tfrac{1}{48}(p_2 - \lambda^2)$.
   From a cohomological point of view, properly describing the action and the corresponding 
 partition function (or path integral) is involved due to the presence of subtle torsion
(see \cite{FM}).  However, we will again evade such subtle issues when we rationalize. 
Having already the interpretation of $[C_3]$ as a String class $\mathcal{S}$, 
 we now interpret  $[\tfrac{1}{6}G_4 \wedge G_4 - I_8]$ as a rational 
cohomology class $y_8 \in H^8(Y^{11}; \Q)$, rationalizing the interpretation at 
the integral level in \cite{DFM}\cite{SSS2}\cite{SSS3}. 
Consequently, we have that the Lagrangian, again at the level of rational cohomology,
 is a variation of a rational Spin-Ninebrane class 
\(
\mathcal{L}_{{}_{M}}^{\rm coh}= \mathcal{N}_\Q- \mathcal{N}'_\Q\;,
\)
where we identify $y_8$ with $\pi^*_{\rm Spin} \psi_8$, and where 
we take $\pi^*_{\rm Spin}\phi_4$ to be zero. 
By dimension reasons, we now automatically have a Ninebrane 
structure on $Y^{11}$, and so the action functional 
captures the trivialization of that structure in the form 
of a Spin-Ninebrane structure. This is similar to having a String structure on the 
M2-brane by virtue of dimension \cite{tcu}.

\medskip
The next two examples deal with the Green-Schwarz anomaly cancellation, which is 
 one of the main highlights of string theory \cite{GS}. 
 We  will first consider the usual mechanism  where we do the matching with 
 expression \eqref{def-nn} using 
  $\pi^*_{\rm Spin} \phi_4=0$, 
  and then consider 
the dual to the Green-Schwarz formulation 
  (in the sense of \cite{Fr}\cite{SSS3}) by taking $\pi^*_{\rm Spin} \psi_8=0$.
   
   \medskip
{\bf Example 3: The Green-Schwarz anomaly cancellation mechanism and variation of 
Spin-Ninebrane structures}.
 This anomaly cancellation arises in heterotic string theory (or type I supergravity).
 The system involves a ten-dimensional manifold $M^{10}$ 
with its natural Spin bundle, a Yang-Mills bundle $E$, a closed 
3-form $H_3$, with corresponding cohomology class $[H_3]$. 
The bundle $E$ enters the expressions via its Chern character ${\rm ch}(E)$, 
while the natural bundles are accounted for via the Pontrjagin classes $p_i(TM^{10})$. 
 The corresponding action functional includes the term
\(
\mathcal{L}_{{}_{\rm GS}}= H_3 \wedge J_8\;,
\)
where $J_8$ is a closed 8-form with cohomology class 
$-{\rm ch}_4(E) + \tfrac{1}{48} p_1(M) {\rm ch}_2(E) - \tfrac{1}{64}p_1(M)^2
+ \tfrac{1}{48} p_2(TM)$. 
Identifying the class $[H_3]$ with the String class $\mathcal{S}$, 
the expression at the level of cohomology is of the form 
$
\mathcal{L}^{\rm coh}_{{}_{\rm GS}}= \mathcal{S} \cdot \psi_8
$,
where we have also identified the rational cohomology class $\psi_8$ with $[J_8]$. 
This gives a special instance of a variation of rational Spin-Ninebrane
structures 
\(
\mathcal{L}^{\rm coh}_{{}_{\rm GS}}=\mathcal{N}_\Q - \mathcal{N}'_\Q\;.
\)
Note that  considerable constraints would be needed in order to ensure that $[J_8]$
 is an integral class (see \cite{SSS2} for a discussion on when this is the case). 
 A virtue of working rationally is also highlighted here in the evasion of these complications.

\medskip
{\bf Example 4: The dual Green-Schwarz anomaly and variation of Spin-Fivebrane 
structures}.
The dual Green-Schwarz anomaly cancellation arises in heterotic 
string theory (or type I supergravity).
The system still involves a ten-dimensional manifold $M^{10}$ 
with its natural Spin bundle, a Yang-Mills bundle $E$, except that now we have the
(Hodge-dual) closed 
form $H_7$, with corresponding cohomology class $[H_7]$, of degree seven.
The action functional involves a term of the form
\(
\mathcal{L}_{{}_{\rm{dual\; GS}}} =H_7 \wedge J_4\;,
\)
where $J_4$ is a 4-form with cohomology class $p_1(TM^{10}) - {\rm ch}_2(E)$. 
Then the cohomology class corresponding to this action functional is given as the difference
\(
\mathcal{L}_{{}_{\rm{dual\; GS}}}^{\rm coh}=\mathcal{N}_\Q - \mathcal{N}'_\Q\;,
\)
where we identify $[H_7]$ with the Spin-Fivebrane class $\mathcal{F}$, 
$[J_4]$ with the form $\pi^*_{\rm Spin} \phi_4$, and where we take
$\pi^*_{\rm Spin}\psi_8$ to be zero.

Rationally, there is an isomorphism
$H^{12}(B\Spin;\Q)/(p_1^\Q, p_2^\Q)\cong H^{12}(B\Five;\Q)$. 
 If $H^4(M;\Z)$ and $H^8(M; \Z)$ are pure torsion, then the set of Ninebrane classes 
 and Spin-Ninebrane classes coincide. 

\medskip
{\bf Example:} We can give a nontrivial example of a spacetime manifold $X$ which has torsion $H^4(X; \Z)$,
vanishing $H^8(X; \Z)$ and non-torsion $H^{12}(X; \Z)$. As in  \cite{Sing},
let ${\rm SU}(2)$ be the subgroup of ${\rm SU}(4)$ consisting of all block diagonal matrices 
${\rm diag}(A, A)$ where $A \in {\rm SU}(2)$. Then the 12-dimensional 
quotient $X={\rm SU}(4)/{\rm SU}(2)$, viewed as 
the base of an $S^3$ bundle,  is stably parallelizable with $H^4(X; \Z)=\Z_2$, 
$H^8(X; \Z)=0$ and $H^{12}(X; \Z)=\Z$.

\medskip
\paragraph{Remarks} The above formulation extends to other theories as well:

\vspace{-2mm}
\begin{enumerate}[{\bf (i)}]
\item   Another string theory, namely type I, 
admits a duality-symmetric formulation, with the dual formulation using 
$H_7$ given in \cite{Cham}.  A similar connection to the above higher 
structures holds in this case. 

\vspace{-2mm}
\item It is also possible to consider duality-symmetric 
M-theory action \cite{BBS} by supplementing the dual (see \cite{Fre})
 to the action to 
\eqref{M action}
where we would have both $\pi_{\rm Spin}^* \psi_4$ and 
$\pi_{\rm Spin}^*\phi_8$ nonzero in general. Hence this would involve 
variations of both Spin-Fivebrane 
and Spin-Ninebrane structures. 

\vspace{-2mm}
\item  Similarly, 
the duality-symmetric 
heterotic action \cite{GN}\cite{Fr} (by combining Examples 3
and 4
would involve both $\pi_{\rm Spin}^* \psi_4$ and 
$\pi_{\rm Spin}^*\phi_8$, hence also leading to 
variation of both structures.

\vspace{-2mm}
\item With appropriate interpretation and conditions on classes, other theories such as 
topological (super)gravity \cite{Ch}, Chern-Simons (super)gravity \cite{Z},
and higher Chern-Simons theories \cite{BG}\cite{FSS}
can also be made to fit the above description. 
\end{enumerate}

We find it noteworthy that the topological parts of  the three main action functionals 
in M-theory, i.e., that of the M2-brane, the M5/NS5-brane, and of classical M-theory,
as well as the Green-Schwarz anomaly and its dual, can be interpreted as trivializations 
of higher obstructions with the rational Spin-Fivebrane and Spin-Ninebrane structures
able to account for the form as well as the compositeness of the actions. The above discussion 
can be summarized in the following schematic table, where the last column displays the 
new interpretation as a variation of a higher structure. 
\begin{center}
{\small
\begin{tabular}{|c||c|c|}
\hline 
{\bf System} & {\bf Existing 
structure} & {\bf Variation of}
\\
\hline
\hline 
Chern-Simons  & Riemannian &   Spin 
\\
\hline 
M2-brane    & Spin &   String   
\\
\hline 
M5/NS5-brane   & String &   Spin-Fivebrane   
\\
\hline 
M-theory, (dual) GS & String/Fivebrane &  Spin-Ninebrane 
\\
\hline 
\end{tabular} 
}
\end{center}

\noindent The first row follows \cite{J}\cite{tcu}\cite{Red}, while 
the other three start with existing descriptions, as presented in 
\cite{SSS2}\cite{tcu}\cite{SSS3}\cite{9brane}, and 
then provides an alternative description,
 as given in the above examples. More explicitly, 
 the new description for the NS5/M5-brane theory is given 
 in Example 1
 that for  M-theory  in Example 2,
 as well as the Green-Schwarz anomaly and its
 dual are given in Examples 3
 and 4,
 respectively.

 What we provided above are only glimpses of connections, and we believe that 
 this is the starting point of interesting constructions, which deserve to be elaborated 
 on in a lot of detail. In particular, a deeper discussion on variations 
 of the structures would have to be in the context of partition functions, rather than just
 action functionals. This would be considerable, as it requires the two notions that we 
 suppressed, namely torsion and higher bundles, and would go way beyond this note, 
 but we plan to take it up elsewhere.

\vspace{.5cm}
\noindent {\large \bf Acknowledgement.} 
The authors thank Urs Schreiber and the anonymous referees of this note and 
of \cite{SW1} for useful comments and suggestions.  


\end{document}